\documentclass[
 reprint,
 amsmath,amssymb,superscriptaddress,
 aps,
 prl,
groupedaddress
]{revtex4-1}

\usepackage{graphicx}
\usepackage{dcolumn}
\usepackage{bm}
\usepackage{color}
\usepackage{ulem}
\usepackage{float}
\usepackage{textcomp}
\restylefloat{table}
\usepackage{verbatim}

\begin{document}

\title{Composite Fermion Geometric Resonance Near $\nu=1/2$ Fractional Quantum Hall State}
\date{\today}

\author{M. A.\ Mueed}
\author{D.\ Kamburov}
\author{S.\ Hasdemir}
\author{M.\ Shayegan}
\author{L. N.\ Pfeiffer}
\author{K. W.\ West}
\author{K. W.\ Baldwin}
\affiliation{Department of Electrical Engineering, Princeton University, Princeton, New Jersey 08544, USA}

\begin{abstract}
We observe geometric resonance features of composite fermions on the flanks of the even-denominator $\nu=1/2$ fractional quantum Hall state in high-mobility two-dimensional electron and hole systems confined to wide GaAs quantum wells and subjected to a weak, strain-induced, unidirectional periodic potential modulation. The features provide a measure of how close to $\nu=1/2$ the system stays single-component and supports a composite fermion Fermi sea before transitioning into a $\nu=1/2$ fractional quantum Hall state, presumably the two-component $\Psi_{331}$ state. 
\end{abstract}

\maketitle

Clean two-dimensional electron systems (2DESs) subjected to a perpendicular magnetic field $B$ exhibit an array of novel phases. At and near $B=0$, the 2D electrons are typically a degenerate Fermi gas and occupy a Fermi sea in their ground-state. At higher $B$, however, their ground-state can transform into various phases such as the integral and fractional quantum Hall states (FQHSs), Fermi gas of composite fermions (CFs), stripe and bubble phases, and Wigner crystal \cite{Jain.2007, Shayegan.Flatland.Review.2006}. The evolution of the system into different phases as a function of $B$, and in some cases temperature, makes the physics of 2DES very exciting. Here we address a fundamental question: Can the 2DES very near Landau level filling factor $\nu=1/2$ be described by CFs possessing a well-defined Fermi surface, if the ground-state at $\nu=1/2$ is an incompressible FQHS?

In a standard, single-layer 2DES, the ground-state at $\nu=1/2$ is known to be a compressible state which is elegantly described by CFs, exotic quasi-particles each composed of an electron and two magnetic flux quanta \cite{Jain.2007,Shayegan.Flatland.Review.2006,Jain.PRL.1989,Halperin.PRB.1993}. One of the fundamental properties of CFs is that they form a Fermi gas and occupy a Fermi sea at $\nu=1/2$ \cite{Halperin.PRB.1993}. This Fermi gas state of CFs, similar to that of electrons near zero $B$, also extends to the vicinity of $\nu=1/2$ where the CFs experience a small effective magnetic field $B^{*}=B-B_{1/2}$ \cite{footnote1}, $B_{1/2}$ being the field at $\nu=1/2$. The existence of a Fermi surface of CFs near $\nu=1/2$ has indeed been confirmed in numerous commensurability experiments where the geometric resonance of the CF cyclotron orbit, defined by $B^{*}$, with laterally-induced sample features were exploited \cite{Willett.PRL.1993, Kang.PRL.1993, Goldman.PRL.1994, Smet.PRL.1996, Smet.PRB.1997, Smet.PRL.1998, Mirlin.PRL.1998, Oppen.PRL.1998, Smet.PRL.1999, Zwerschke.PRL.1999, Willett.PRL.1999, Kamburov.PRL.2012, Kamburov.PRL.2013, Kamburov.PRB.2014, Kamburov.PRL.2014}. Now, under suitable conditions, namely in wide quantum wells (QWs) with bilayer-like charge distribution \cite{Suen.PRL.1992,Suen1.PRL.1992,Suen.PRL.1994,Shabani.PRB.2013}, the 2DES can prefer FQHS as the ground-state at $\nu=1/2$ over the Fermi gas of CFs. The commensurability experiments we report here for such systems reveal that CFs with a Fermi sea exist near $\nu=1/2$ even though the ground-state at $\nu=1/2$ is no longer a Fermi gas but an incompressible FQHS.

The 2DES sample we studied is confined to a 65-nm-wide symmetric GaAs QW, grown by molecular beam epitaxy on a (001) GaAs substrate. The QW, located 190 nm below the surface, is flanked on each side by 95-nm-thick Al$_{0.24}$Ga$_{0.76}$As spacer layers and Si $\delta$-doped layers. The 2DES density is $\simeq 1.8\times10^{11}$ cm$^{-2}$, and mobility $\simeq10^{7}$ cm$^{2}$/Vs. As shown in Fig. 1, we pattern the surface of a standard Hall bar with a strain-inducing superlattice of period $a=200$ nm. The superlattice, made of negative electron-beam resist, produces a potential modulation of the same period in the 2DES through the piezoelectric effect in GaAs \cite{Kamburov1.PRB.2012, Kamburov2.PRB.2012, Skuras.APL.1997, Endo.PRB.2000, Endo.PRB.2001, Kamburov.PRL.2012, Kamburov.PRB.2014, Kamburov.PRL.2013, Kamburov.PRL.2014}. Part of the Hall bar is intentionally left unpatterned as a reference region. We pass current along the Hall bar and simultaneously measure the magnetoresistance of the patterned and the reference regions. The potential modulation in the patterned region induces commensurability features stemming from the geometric resonance of CFs with the period of modulation. Using these features, we probe the Fermi surface properties of CFs in the vicinity of the $\nu=1/2$ FQHS. Measurements are carried out using a dilution refrigerator or a $^{3}He$ cryostat.

\begin{figure}
\includegraphics[width=.46\textwidth]{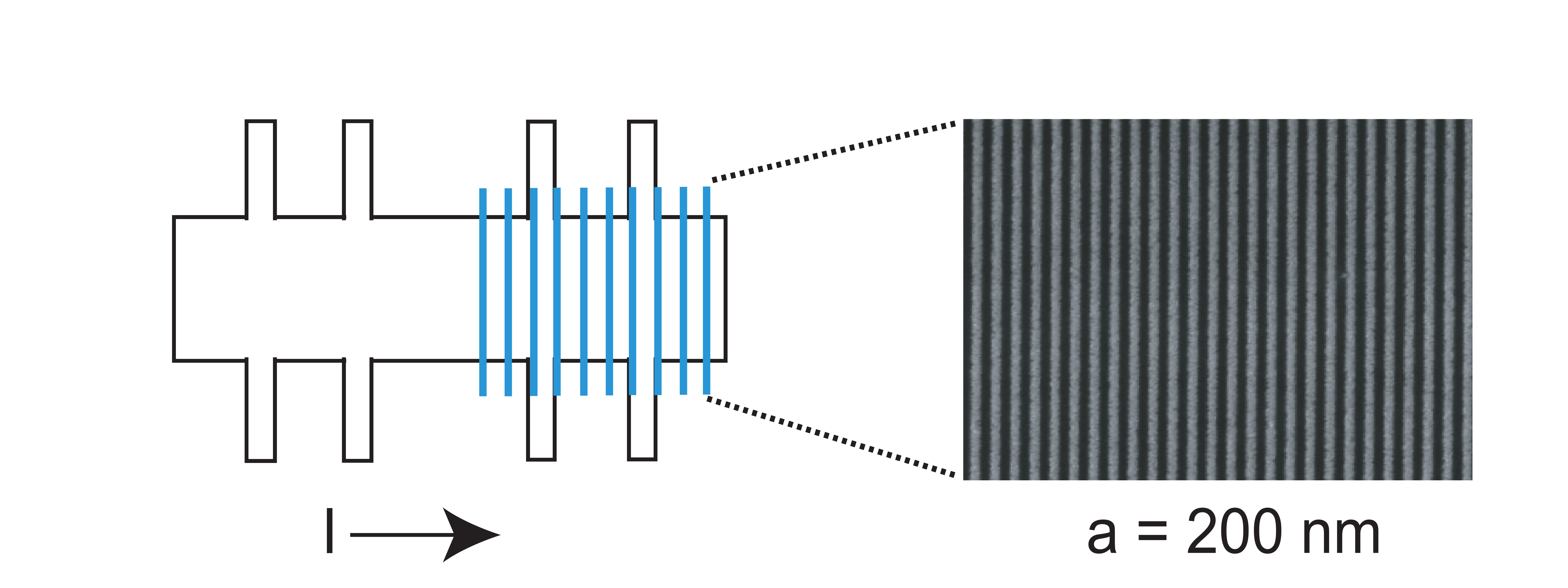}
\caption{\label{fig:Fig1} (Color online) (a) Sample schematic. The electron beam resist grating covering part of the Hall bar surface is represented as blue stripes; the unpatterned region on the left serves as a ``reference". The scanning electron microscope image shows the surface pattern from the patterned region.}
\end{figure}

\begin{figure*}
\includegraphics[width=1\textwidth]{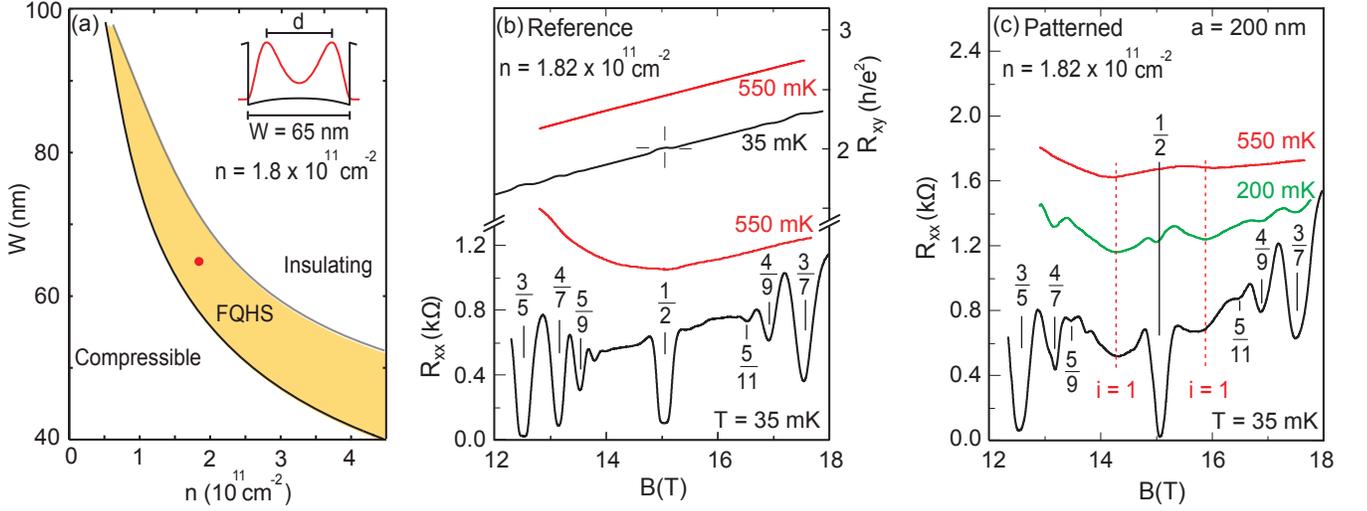}
\caption{\label{fig:Fig1} (Color online) (a) The well width ($W$) vs density ($n$) phase diagram for the different ground-states of the 2DES at $\nu=1/2$ in symmetric, wide GaAs QWs \cite{Shabani.PRB.2013}. The yellow region is where FQHS is observed. The ground-state changes to ``compressible" and ``insulating" to the left and right of the FQHS region, respectively. The red circle marks the position of our 2DES sample in the phase diagram. Inset: Results of self-consistent calculations of the charge distribution and Hartree potential for our 2DES sample at $B=0$. (b) Longitudinal ($R_{xx}$) and Hall ($R_{xy}$) magnetoresistance traces from the reference (\textit{unpatterned}) region of the sample. (c) $R_{xx}$ traces from the \textit{patterned} region of the sample. In both (b) and (c) traces taken at different temperatures are shifted vertically for clarity. }
\end{figure*}

The phase diagram of Fig. 2(a) illustrates the experimental conditions under which the $\nu=1/2$ FQHS is observed in 2DESs confined to symmetric, wide GaAs QWs \cite{Suen.PRL.1994,Shabani.PRB.2013}. An important property of wide QWs is that both $\Delta_{SAS}$ (symmetric-to-antisymmetric subband separation) and $d$ (the inter-layer separation, see Fig. 2(a) inset), which characterize the coupling between the layers, can be tuned by changing the 2DES density ($n$) \cite{Suen.PRB.1991,Suen.PRL.1992,Suen1.PRL.1992,Suen.PRL.1994,Shabani.PRB.2013}. For a given well width ($W$), at low $n$, when $\Delta_{SAS}$ is large, the ground-state at $\nu=1/2$ is compressible. Increasing $n$ makes $\Delta_{SAS}$ smaller and $d$ larger so that the charge distribution becomes increasingly bilayer-like and a $\nu=1/2$ FQHS is stabilized, presumably as a two-component ($\Psi_{331}$) state. At higher $n$, however, the ground-state at $\nu=1/2$ turns insulating, likely due to the formation of a bilayer Wigner crystal \cite{Manoharan.PRL.1996, Hatke.arxiv.2015}. The yellow region of Fig. 2(a) phase diagram illustrates the range of $W$ and $n$ required to stabilize the $\nu=1/2$ FQHS \cite{Shabani.PRB.2013}. We chose the values of $W$ and $n$ for our 2DES sample near the center of this region. The traces from the $reference$ region show a deep minimum in longitudinal magnetoresistance ($R_{xx}$) and a plateau at $2h/e^{2}$ in the Hall resistance ($R_{xy})$; see the 35 mK traces in Fig. 2(b). These are clear indications for a well-developed FQHS at $\nu=1/2$ for the 2DES sample, consistent with its position in the phase diagram. As seen in the 550 mK traces, the minimum and plateau at $\nu=1/2$ vanish at higher temperature. 

Before discussing the data from the $patterned$ region, we briefly describe the theoretical picture of the commensurability features for $\nu=1/2$ CFs in standard (single-layer) 2DESs where the ground-state at $\nu=1/2$ is compressible \cite{Willett.PRL.1999, Smet.PRB.1997, Smet.PRL.1998, Mirlin.PRL.1998, Oppen.PRL.1998, Smet.PRL.1999, Zwerschke.PRL.1999, Kamburov.PRL.2013, Kamburov.PRL.2014, Kamburov.PRL.2012, Kamburov.PRB.2014}. These features are manifested as $R_{xx}$ minima near $\nu=1/2$ whenever the quasi-classical cyclotron orbit diameter $2R_{c}^{*}$ of CFs becomes commensurate with integer multiples of the density modulation period $a$. More precisely, resistance minima are observed at the magnetic commensurability condition $2R_{c}^{*}/a=i+1/4$ \cite{Willett.PRL.1999, Smet.PRB.1997, Smet.PRL.1998, Mirlin.PRL.1998, Oppen.PRL.1998, Smet.PRL.1999, Zwerschke.PRL.1999, Kamburov.PRL.2013, Kamburov.PRL.2014, Kamburov.PRL.2012, Kamburov.PRB.2014}, where i = 1, 2, 3..., $2R_{c}^{*}={2\hbar}k_{F}^{*}/eB^{*}$, $B^{*}=B-B_{1/2}$ is the effective magnetic field seen by the CFs, and $k_{F}^{*}={\sqrt{4{\pi}n^{*}}}$ is the Fermi wave vector of CFs \cite{footnote2}. The expression for $k_{F}^{*}$ assumes full spin polarization of CFs since $\nu=1/2$ is typically positioned at very high fields. In our study, we focus on the commensurability resistance minima on the sides of $\nu=1/2$ corresponding to $i=1$. These minima give a direct measure of $k_{F}^{*}$ thus providing a clear signature for the existence of a well-defined CF Fermi surface. 

\begin{figure*}
\includegraphics[width=1\textwidth]{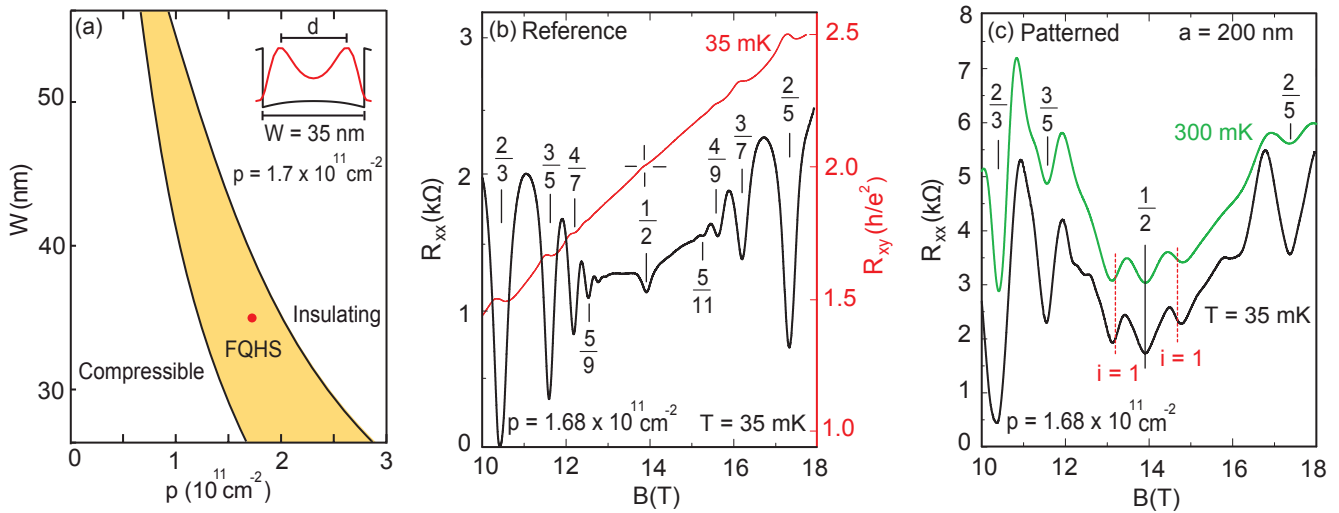}
\caption{\label{fig:Fig1} (Color online) (a) Phase diagram for the different ground-states at $\nu=1/2$ of a 2DHS confined to symmetric, wide GaAs QWs \cite{Liu.PRL.2014}. The red circle marks the position of our 2DHS sample in the phase diagram. Inset: Self-consistently calculated charge distribution and Hartree potential for the 2DHS sample at $B=0$. (c) $R_{xx}$ and $R_{xy}$ traces from the \textit{reference} region of the sample. (c) $R_{xx}$ traces from the \textit{patterned} region. Traces at different temperatures are shifted vertically for clarity.}
\end{figure*}

Figure 2(c) is the highlight of our study. It shows three $R_{xx}$ traces from the patterned region of the 2DES sample, taken at different temperatures. Similar to the reference region, we observe a deep minimum at $\nu=1/2$ in the $T=35$ mK trace indicating a strong FQHS. Remarkably, in addition to the FQHS minimum at $\nu=1/2$, there are two minima on the sides of $\nu=1/2$, flanked by shoulders of increasing resistance. We also mark, with dotted red lines on both sides of $\nu=1/2$, the expected field-positions of the commensurability minima based on the commensurability condition $2R_{C}^{*}/a=5/4$ and the total 2DES density. These lines agree very well with the positions of the observed minima. The absence of such minima near $\nu=1/2$ FQHS in the 35 mK trace from the reference region (bottom trace of Fig. 2(b)) is very suggestive that the minima observed in Fig. 2(c) indeed correspond to the commensurability of CFs' cyclotron orbit with the periodic modulation. The temperature dependence of $R_{xx}$ in Fig. 2(c) also supports our conjecture. In the trace taken at $T=200$ mK, the strength of the FQHS minima at $\nu=1/2, 4/7, 4/9, 3/7$ significantly weakens, whereas the two minima near $\nu=1/2$ are clearly less affected. At $T=550$ mK, the aforementioned FQHSs completely disappear. However, we still observe signs of weak minima matching the dotted red lines. Considering the fact that commensurability features, which originate from the quasi-classical CF cyclotron orbits, are less temperature dependent than the FQHSs, we conclude that the observed minima here signal the Fermi surface of CFs. In summary, the ground-state \textit{near} $\nu=1/2$ in our 2DES is clearly a Fermi gas of CFs while at $\nu=1/2$ it is an incompressible FQHS.

We emphasize that in standard, single-layer 2DES the ground-state at and near $\nu=1/2$ is compressible and is described by a Fermi sea of CFs. Our findings establish that a Fermi sea of CFs can exist very near $\nu=1/2$ even though the ground-state at $\nu=1/2$ is no longer a Fermi gas state of CFs. This also highlights an important difference between the properties of electrons and CFs. In the case of degenerate electrons, the ground-state at $B=0$ is compressible with a well-defined Fermi surface which is also preserved in the limit of small $B$. Commensurability features for electrons in small $B$ thus signal the existence of a degenerate Fermi sea at $B=0$. Furthermore, such features provide a measure of the Fermi wave vector at $B=0$. In contrast, as illustrated by the data of Fig. 2(c), the absence of a CF Fermi sea at $B^{*}=0$ does not rule out the existence of a CF Fermi surface at small $B^{*}$.

As stated earlier, the $\nu=1/2$ FQHS observed in 2DESs confined to symmetric, wide, single QWs is generally believed to be the two-component $\Psi_{331}$ state \cite{Shabani.PRB.2013}. However, other FQHSs neighboring $\nu=1/2$ resemble those only observed in 2DESs with standard single-layer charge distributions; examples include the odd-numerator states such as those at $\nu=5/9$ and 5/11 which are clearly single-component. This suggests that the system undergoes a single-component to two-component phase transition near $\nu=1/2$. However, in the very close vicinity of $\nu=1/2$ where no FQHSs are observed, it is unclear whether the system stays single-component. The commensurability features which we observe in this particular region clearly demonstrate that the system does remain single-component very close to $\nu=1/2$ and then undergoes a sharp phase transition into a two-component, imcompressible state at $\nu=1/2$. It is particularly noteworthy that our observed commensurability minima in Fig. 2(c) are consistent with the \textit{total} density of the 2DES. 

Next we focus on Fig. 3 which shows similar results for a 2D \textit{hole} system (2DHS). In this sample, the 2D holes are confined to a 35-nm-wide symmetric GaAs QW, grown by molecular beam epitaxy on a (001) GaAs substrate. The QW located 135 nm below the surface is flanked on each side by 95-nm-thick Al$_{0.24}$Ga$_{0.76}$As spacer layers and C $\delta$-doped layers. The 2DHS density ($p$) is $\simeq 1.68\times10^{11}$ cm$^{-2}$, and mobility is $\simeq10^{6}$ cm$^{2}$/Vs. The phase diagram describing different ground-states for $\nu=1/2$ in 2DHSs is shown in Fig. 3(a) \cite{Liu.PRL.2014}. This phase diagram differs from the one for 2DESs (Fig. 2(a)) in that, for a given $W$, the $\nu=1/2$ FQHS is observed at a smaller density. The basic reason for this difference is that the larger hole effective mass renders the 2DHS into a bilayer-like system at a smaller density \cite{Liu.PRL.2014}. The $W$ and $p$ for our 2DHS sample indicate that, analogous to the 2DES case, the ground-state at $\nu=1/2$ should be a FQHS. This is indeed the case as we observe a clear $R_{xx}$ minimum at $\nu=1/2$ and a developing $R_{xy}$ plateau at $2h/e^{2}$ in Fig. 3(b). However, compared to the 2DES sample, the FQHS at $\nu=1/2$ is considerably weaker partly because of the lower mobility of the 2DHS sample \cite{footnoteA}. 

The $R_{xx}$ data of Fig. 3(c) from the patterned region at $T=35$ mK show signatures of the commensurability features near $\nu=1/2$. The minima are slightly further away from $\nu=1/2$ than the dotted red lines which are based on the total 2DHS density. The positions of these minima are consistent with the $warped$ geometry of CFs' Fermi surface in this 2DHS sample as discussed elsewhere \cite{Mueed.arxiv.2015}. Similar to the reference region, the minimum at $\nu=1/2$ is not strong. However, the $R_{xx}$ trace taken at 300 mK shows a weakening of the $\nu=1/2$ minimum relative to the commensurability features. We conclude that similar to the 2DES case, an incompressible FQHS at $\nu=1/2$ is flanked by a Fermi sea of CFs very near $\nu=1/2$.

To summarize, results from both 2DES and 2DHS samples show that the system is able to retain a well-defined Fermi surface of CFs in close proximity of $\nu=1/2$ even though the ground-state at $\nu=1/2$ is not a Fermi gas but an incompressible FQHS. We would like to emphasize that the technique used in our work can also be applied to explore certain long-standing puzzles about the enigmatic FQHS at $\nu=5/2$. It is widely believed that the pairing of CFs leads to the formation of $\nu=5/2$ FQHS \cite{Willett.Rep.2013}. Fermi surface properties of CFs near $\nu=5/2$ were indeed explored previously through surface acoustic wave measurements \cite{Willett.Rep.2013,Willett.PRL.2002}. While an enhanced conductivity near $\nu=5/2$ at high temperatures hinted at the existence of a Fermi surface, no geometric resonance was observed on the flanks of $\nu=5/2$ \cite{Willett.PRL.2002}. Inducing commensurability minima near the $\nu=5/2$ FQHS would be a more direct demonstration of a Fermi surface. In addition, the Fermi wave vector of CFs deduced from the positions of these minima could shed light on whether the $\nu=5/2$ FQHS is two-component (spin-unpolarized) or single-component (spin-polarized).

We acknowledge support through the NSF (Grants ECCS-1508925 and MRSEC DMR-1420541), the Gordon and Betty Moore Foundation (Grant GBMF4420), and the Keck Foundation for sample fabrication and characterization, and the DOE BES (Grant DE-FG02-00-ER45841) and the NSF (Grant DMR-1305691) for measurements on 2DHSs and 2DESs, respectively. Our work was performed at the National High Magnetic Field Laboratory, which is supported by the NSF Cooperative Agreement DMR-1157490, by the State of Florida, and by the DOE. We thank Yang Liu and R. Winkler for illuminating discussions and providing the insets of Figs. 2(a) and 3(a), respectively. We also thank S. Hannahs, T. Murphy, A. Suslov, J. Park and G. Jones at NHMFL for technical support.


\begin{thebibliography}{99}
\bibitem{footnote1}Throughout this paper, we use `` * " to denote the CF parameters.
\end{thebibliography}
\end{document}